# Generic and Typical Ranks of Three-Way Arrays

P. Comon[1], J. ten Berge[2]

(1) Lab. I3S, B.P.121, F-06903 Sophia-Antipolis cedex, France

(2) University of Groningen, 9712 TS Groningen, The Netherlands

**Abstract**

The concept of tensor rank, introduced in the twenties, has been popularized at the beginning of the seventies. This has allowed to carry out Factor Analysis on arrays with more than two indices. The generic rank may be seen as an upper bound to the number of factors that can be extracted from a given tensor. We explain in this short paper how to obtain numerically the generic rank of tensors of arbitrary dimensions, and compare it with the rare algebraic results already known at order three. In particular, we examine the cases of symmetric tensors, tensors with symmetric matrix slices, or tensors with free entries.

**Résumé**

La notion de rang tensoriel, proposée dans les années vingt, a été popularisée au début des années soixante-dix. Ceci a permis de mettre en oeuvre l'Analyse de Facteurs sur des tableaux de données comportant plus de deux indices. Le rang générique peut être vu comme une borne supérieure sur le nombre de facteurs pouvant être extraits d'un tenseur donné. Nous expliquons dans ce court article comment trouver numériquement le rang générique d'un tenseur de dimensions arbitraires, et le comparons aux quelques rares résultats algébriques déjà connus à l'ordre trois. Nous examinons notamment les cas des tenseurs symétriques, des tenseurs à tranches matricielles symétriques, ou des tenseurs à éléments libres.

**Acknowledgment :** This work has been presented in part at the TRICAP Workshop, Chania, Crete, Greece, June 4-9, 2006. It has been partly supported by the IST Programme of the European Community, under the PASCAL Network of Excellence, IST-2002-506778. This publication only reflects the authors' views.

# 1 Introduction

Generic ranks, defined in the complex field, have been studied for several decades [10] [14]. However, the value of the generic rank for arbitrary dimensions is not yet known in the unsymmetric case, and has been known in the symmetric case only recently [4]. The existence itself of the generic rank is not ensured in the real case, and there exist in general several *typical ranks* (see section 2.1 for definitions). The typical tensorial rank of three-way arrays over algebraically closed fields has been much studied in the context of computational complexity theory. Bürgisser, Clausen and Shokrollahi [2, Ch.20] give an overview of general results for various classes of arrays; these results have been extended in [4]. The study of tensorial rank over the real field has lagged behind. In this paper, generic and typical ranks are discussed for various tensor structures.

The typical rank of three-way arrays over the real field has been relevant for psychological data analysis since Carroll and Chang [3] and Harshman [9] independently proposed a method which they christened CANDECOMP and PARAFAC, respectively. This *Canonical Decomposition* (CAND) method generalizes Principal Component Analysis to three-way data, by seeking the best least squares approximation of the data array by the sum of a limited number of rank-one arrays. In 2-way analysis, the rank of the data matrix is the maximum number of components that Principal Component Analysis can extract. This property generalizes smoothly to three-way data. That is, the rank of a three-way array is the maximum number of components that CAND can extract. Thus, the study of typical rank of three-way arrays is of great theoretical importance for CAND.

Although CAND was developed in a psychometric environment, its main area of applications has been Chemometrics, e.g. [13]. In addition to straightforward application of CAND, chemometricians also use Tucker3 component analysis [21] quite often. This is a generalized version of CAND which decomposes a data array as weighted sum of rank-one arrays, the weights being collected in a so-called core array. Typically, the underlying chemometric model dictates that a vast majority of specified core elements vanish. Because there exist admissible transformations which generate a vast majority of zero elements in arbitrary arrays, we need tools to tell models from tautologies. This is were the concept of typical rank has found another realm of application. For instance, Ten Berge and Smilde [19] have argued that a sparse core hypothesized by Gurden et alterae [7] is indeed a model and not a tautology. Their hypothetical core was a $5 \times 5 \times 3$ array with only 5 nonzero entries, hence of rank 5 at most. Because $5 \times 5 \times 3$ arrays have a typical rank of at least 7, it is clear at once that transformations which yield as few as five non-zero elements, starting from any randomly generated $5 \times 5 \times 3$ array, do not exist. In this way, the typical rank of three-way arrays find applications in distinguishing constrained Tucker3 models from tautologies.

The paper is organized as follows. In section 2, definitions and historical remarks are provided. Next, a numerical algorithm is described in section 3, which is able to compute the generic rank of any tensor, symmetric or not. Numerical values are reported in section 4, and compared to the already known rank values previously obtained by means of algebraic calculations. The consistency of the results confirm the validity of the approach, which can yield generic ranks for more complicated structures, such as tensors with symmetric matrix slices (sometimes referred to as the INDSCAL model), among others.

# 2 Generic and Typical Ranks

## 2.1 Definitions

Let $\boldsymbol{T}$ be a $L$-way array of dimensions $N_\ell$, $1 \leq \ell \leq L$, with values in a ring $\mathcal{R}$. This array always admits a decomposition into a sum of outer products as:

$$\boldsymbol{T} = \sum_{r=1}^{R} \boldsymbol{u}_r^{(1)} \otimes \boldsymbol{u}_r^{(2)} \otimes \ldots \otimes \boldsymbol{u}_r^{(L)} \qquad (1)$$

where $\boldsymbol{u}_r^{(\ell)}$ is a $N_\ell \times 1$ array, $\forall r$, and $\otimes$ denotes the tensor product.

Now consider an array $\boldsymbol{T}$ with values in a field $\mathbb{K}$. Arrays $\boldsymbol{u}_r^{(\ell)}$ may be considered as vectors of the linear space $\mathbb{K}^{N_\ell}$. Thus, as a combination of tensor products of vectors, $\boldsymbol{T}$ may be considered as a tensor. Under a linear change of coordinate system in each space $\mathbb{K}^{N_\ell}$, defined by a matrix $\boldsymbol{A}^{(\ell)}$, the tensor is represented by another array, obtained by the multi-linear transform $\{\boldsymbol{A}^{(1)}, \boldsymbol{A}^{(2)}, \ldots \boldsymbol{A}^{(L)}\}$. Since it is legitimate once a basis has been defined in the space, no distinction will be made in the remainder between the tensor and its array representation.

The *rank* of a given tensor $\boldsymbol{T}$ (and by extension, of the array defining its coordinates in a given basis) is the minimal integer $R$ such that the decomposition (1) is exactly satisfied. This decomposition is referred to as the tensor Canonical Decomposition (CAND).

A property is called *typical* if it holds true on a set of nonzero volume [4]. This supposes that some topology has been defined on $\mathbb{K}^{N_1 \times N_2 \times \ldots N_L}$; this can be the Zariski topology for instance, or an Euclidian topology. A property is said to be *generic* if it is true almost everywhere. In other words, a generic property is typical, but the converse is not true.

Let $N_\ell$ be $L$ given positive integers. Then the rank of tensors of size $N_1 \times N_2 \times \cdots \times N_l$ is bounded, and one can make a partition of the tensor space, according to the rank values. One can define *typical ranks* as the ranks that are associated with subsets of nonzero volume in the latter partition. If there is a single typical rank, then it may be called the *generic rank*.

For instance, there is a single generic rank if the underlying field $\mathbb{K}$ is algebraically closed (as the field of complex numbers, $\mathbb{C}$) [14] [4]. But there may be several typical ranks if $\mathbb{K}$ is the real field, $\mathbb{R}$.

## 2.2 Historical remarks

The study of typical rank over the real field was initiated by Kruskal [11] [12], who noted that $2 \times 2 \times 2$ arrays had both rank 2 and rank 3 with positive probability. Kruskal also added a few typical ranks for small arrays. Franc [6] discussed some more results, including bounds on typical rank. Ten Berge and Kiers [17] gave a first result of some generality, in solving the typical rank issue for all two-slices arrays (that is, arrays of format $2 \times J \times K$). These results were further generalized in [15], to include all cases where, for $I \geq J \geq K$, $I > JK - J$. Additional miscellaneous results can be found in [1] [15] [16] [20].

When Carroll and Chang developed Candecomp, the main applications they had in mind (a scalar product fitting problem related to INDSCAL) involves three-way arrays with slices that are symmetric in two of the three modes. Ten Berge, Sidiropoulos and Rocci [18] noted that this form of symmetry would affect typical ranks, and examined a number of cases. Quite often, indeed, symmetry of slices appears to entail lower typical rank values. On the other hand, there are also cases where symmetry of the slices does not affect the typical rank. A partial explanation for this can be found in [20]. Ten Berge et al. also noted that symmetric slices are often double centered [3, p. 286], which will further reduce the typical rank. That is, when an array has I double centered slices of order $J \times J$, it can be reduced to a $I \times (J-1) \times (J-1)$ array, and its typical rank will therefore be the same as that of noncentered symmetric $I \times (J-1) \times (J-1)$ arrays. A parallel reasoning can be given for nonsymmetrical double centered slices. A rationale for double centering slices in the PARAFAC context can be found in [8, p.239]. It is easy to show that the $I \times J \times K$ array with I double centered slices has the same typical rank as the uncentered $I \times (J-1) \times (K-1)$ array.

## 3 Computation of Generic Ranks

The algorithm proposed is directly inspired from [5]. Equation (1) can be seen as a parametrization of tensor $\boldsymbol{T}$. In fact, given a set of vectors $\{\boldsymbol{u}_r^{(\ell)} \in \mathbb{K}^{N_\ell}, 1 \leq \ell \leq L, 1 \leq r\}$, consider the mapping $\varphi$ defined from a known subspace $\mathcal{T}_R$ of $(\mathbb{K}^{N_1} \times \mathbb{K}^{N_2} \times \cdots \times \mathbb{K}^{N_L})^R$ onto $\mathbb{K}^{N_1 N_2 \ldots N_L}$ as:

$$\{\boldsymbol{u}_r^{(\ell)} \in \mathcal{T}_R, 1 \leq \ell \leq L, 1 \leq r \leq R\} \to \sum_{r=1}^{R} \boldsymbol{u}_r^{(1)} \otimes \boldsymbol{u}_r^{(2)} \otimes \ldots \otimes \boldsymbol{u}_r^{(L)}$$

Denote $\mathcal{Z}_R = \varphi(\mathcal{T}_R)$ the image of this mapping. Then the dimension $D$ of its closure $\bar{\mathcal{Z}}_R$ is given by the rank of the Jacobian of $\varphi$, expressed in any fixed basis of $\mathbb{K}^{N_1 N_2 \ldots N_L}$. If the Jacobian is of maximal rank, that is, if its rank equals the number $M$ of free parameters in $\mathcal{T}_R$, then it means that $R$ is a typical rank. Actually, $R$ will be either the smallest typical rank, or the generic rank.

Note that it is always possible to reach the maximal Jacobian rank by increasing the number of terms $R$, so that the smallest typical rank is always found.

This result yields the following numerical algorithm.

- Express formally the parametrized rank-one tensor term in a canonical basis
- Express formally the gradient of the latter in this basis
- Draw randomly the parameters according to an absolutely continuous distribution, and initialize matrix $\boldsymbol{J}$ with the numerical value of the gradient
- While rank($\boldsymbol{J}$) strictly increases, do:
  - Draw randomly the parameters according to an absolutely continuous distribution, and append this new numerical value of the gradient as a new row block in $\boldsymbol{J}$
  - Compute the new value of $D = \text{rank}(\boldsymbol{J})$
- Compute the dimension of the fiber of solutions as $F = M - D$, the difference between the number of parameters and the dimension of the image $\bar{\mathcal{Z}}_R$.

In order to clarify the description of this algorithm, we give now the exact expressions of the Jacobian in various cases.

## 3.1 Jacobian for 3rd order asymmetric tensors with free entries

The mapping takes the form below

$$\{\boldsymbol{a}(\ell), \boldsymbol{b}(\ell), \boldsymbol{c}(\ell)\} \xrightarrow{\varphi} \boldsymbol{T} = \sum_{\ell=1}^{r} \boldsymbol{a}(\ell) \otimes \boldsymbol{b}(\ell) \otimes \boldsymbol{c}(\ell)$$

taking into account the presence of redundancies, the number of parameters in this parametrization is $M = R(N_1 + N_2 + N_3 - 2)$. In a canonical basis, $\boldsymbol{T}$ has the coordinate vector:

$$\sum_{\ell=1}^{r} \boldsymbol{a}(\ell) \otimes \boldsymbol{b}(\ell) \otimes \boldsymbol{c}(\ell)$$

where we may decide that $\boldsymbol{a}$, $\boldsymbol{b}$, and $\boldsymbol{c}$ are row arrays of dimension $N_1$, $N_2$, and $N_3$, respectively, and $\otimes$ denotes the Kronecker product. Hence, after $r$ iterations, the Jacobian of $\varphi$ is the $r(N_1 +$

$N_2 + N_3) \times N_1 N_2 N_3$ matrix:

$$J = \begin{bmatrix}
\boldsymbol{I}_{n_1} & \otimes & \boldsymbol{b}^{\mathrm{T}}(1) & \otimes & \boldsymbol{c}^{\mathrm{T}}(1) \\
\boldsymbol{a}(1)^{\mathrm{T}} & \otimes & \boldsymbol{I}_{n_2} & \otimes & \boldsymbol{c}^{\mathrm{T}}(1) \\
\boldsymbol{a}(1)^{\mathrm{T}} & \otimes & \boldsymbol{b}(1)^{\mathrm{T}} & \otimes & \boldsymbol{I}_{n_3} \\
\boldsymbol{I}_{n_1} & \otimes & \ldots & \otimes & \ldots \\
\ldots & \otimes & \boldsymbol{I}_{n_2} & \otimes & \ldots \\
\ldots & \otimes & \ldots & \otimes & \boldsymbol{I}_{n_3} \\
\boldsymbol{I}_{n_1} & \otimes & \boldsymbol{b}^{\mathrm{T}}(r) & \otimes & \boldsymbol{c}^{\mathrm{T}}(r) \\
\boldsymbol{a}(r)^{\mathrm{T}} & \otimes & \boldsymbol{I}_{n_2} & \otimes & \boldsymbol{c}^{\mathrm{T}}(r) \\
\boldsymbol{a}(r)^{\mathrm{T}} & \otimes & \boldsymbol{b}(r)^{\mathrm{T}} & \otimes & \boldsymbol{I}_{n_3}
\end{bmatrix}$$

The values of the generic rank obtained with this algorithm, called `rangj3(N1,N2,N3)`, or `rangj(N,L)` for tensors of arbitrary order $L$ and equal dimensions, are reported in tables 1, 2, and 3.

## 3.2 Jacobian for 3rd order asymmetric tensors with symmetric matrix slices

In this section, consider tensors of size $J \times J \times K$, having symmetric $J \times J$ matrix slices. Our code `rgindscal3(J,K)` implements the computation of the rank of the Jacobian below, when its size increases according to the algorithm described in section 3:

$$J = \begin{bmatrix}
\boldsymbol{I}_J \otimes \boldsymbol{b}(1) \otimes \boldsymbol{c}(1) + \boldsymbol{b}(1) \otimes \boldsymbol{I}_J \otimes \boldsymbol{c}(1) \\
\boldsymbol{b}(1) \otimes \boldsymbol{b}(1) \otimes \boldsymbol{I}_K \\
\ldots \\
\boldsymbol{I}_J \otimes \boldsymbol{b}(r) \otimes \boldsymbol{c} + \boldsymbol{b}(r) \otimes \boldsymbol{I}_J \otimes \boldsymbol{c}(r) \\
\boldsymbol{b}(r) \otimes \boldsymbol{b}(r) \otimes \boldsymbol{I}_K
\end{bmatrix}$$

After $r$ iterations, this matrix is of size $r(J + K) \times J^2 K$. The number of parameters in this parametrization is $M = R(J + K - 1)$. Values of the generic rank are reported in table 4.

## 3.3 Jacobian for 3rd order double centered asymmetric tensors

Now, take again $J \times J \times K$ tensors with symmetric $J \times J$ matrix slices, but assume in addition that every row and column in the latter matrix slices are zero-mean. In order to achieve this, it is sufficient to generate vectors $\boldsymbol{b}(r)$ with zero mean; in otehr words, only $J - 1$ random numbers need to be drawn, the last entry of each vector $\boldsymbol{b}(r)$ being obtained via $b_J = -\sum_{j=1}^{J-1} b_j$. The

Jacobian takes then the expression below:

$$J = \begin{bmatrix} [I_{J-1}, -1] \otimes b(1) \otimes c(1) + b(1) \otimes [I_{J-1}, -1] \otimes c(1) \\ b(1) \otimes b(1) \otimes I_K \\ \ldots \\ [I_{J-1}, -1] \otimes b(r) \otimes c(r) + b(r) \otimes [I_{J-1}, -1] \otimes c(r) \\ b(r) \otimes b(r) \otimes I_K \end{bmatrix}$$

where $\mathbf{1}$ denotes a column of ones of size $J - 1$. At the $r$th iteration, this matrix is of size $r(J + K - 1) \times J^2 K$. The number of parameters in this parametrization is $M = R(J + K - 2)$. Table 5 reports some numerical values obtained with the code `rgindscal2z`.

### 3.4 Jacobian for symmetric tensors

In the case of symmetric tensors of dimension $N$ and order $L$, the mapping $\varphi$ is defined from $\mathbb{K}^{NR}$ to the space of symmetric tensors [5], or equivalently to $\mathbb{K}^p$ with $p = \binom{N+L-1}{L}$, as:

$$\{u(\ell) \in \mathbb{K}^N, \ 1 \leq \ell \leq R\} \xrightarrow{\varphi} \sum_{\ell=1}^{r} u(\ell)^{\otimes L}$$

where $\otimes$ stands for the tensor (outer) product; once a basis is chosen, the tensor product may be replaced by a Kronecker product, yielding exactly the same expression. In the case of order-3 tensors ($L = 3$) and after $r$ iterations, the Jacobian of $\varphi$ takes the following form, somewhat simpler than the previous cases:

$$J = \begin{bmatrix} I_N \otimes a(1) \otimes a(1) + a(1) \otimes I_N \otimes a(1) + a(1) \otimes a(1) \otimes I_N \\ \ldots \\ I_N \otimes a(r) \otimes a(r) + a(r) \otimes I_N \otimes a(r) + a(r) \otimes a(r) \otimes I_N \end{bmatrix}$$

This matrix is of size $rN \times N^3$, but we know that its rank cannot exceed $\binom{N+2}{3} = N(N+1)(N+2)/6$. The number of parameters in this parametrization is $M = RN$. Numerical values of the generic rank obtained with `rangjs(N,L)` are reported in table 6.

## 4 Numerical results

The available results on unconstrained, slicewise symmetric, and double centered arrays can be compared with the numerical values delivered by the computer codes.

**Tensors with free entries.** Table 1 reports typical ranks for 2-slice, 3-slice, and 4-slice arrays. All known typical rank values are reported in plain face, and coincide with the results from `rangj3`. As can be verified, the smallest of the known typical rank values within a cell coincides throughout with the results from `rangj3`. For the unknown entries, we insert the results from `rangj3` in bold face. These bold face values represent the smallest typical rank.

Table 1: Typical ranks for 2-slice, 3-slice, and 4-slice unconstrained arrays. Values reported in bold correspond to generic ranks computed numerically. Values separated by commas are known typical ranks.

|      | K=2 |     |     | K=3 |     |     | K=4 |     |
|------|-----|-----|-----|-----|-----|-----|-----|-----|
|      | J=2 | J=3 | J=4 | J=3 | J=4 | J=5 | J=4 | J=5 |
| I=2  | 2,3 | 3   | 4   | 3,4 | 4   | 5   | 4,5 | 5   |
| I=3  | 3   | 3,4 | 4   | 5   | **5** | 5,6 | **6** | **6** |
| I=4  | 4   | 4   | 4,5 | 5,? | **6** | **6** | **7** | **8** |
| I=5  | 4   | 5   | 5   | 5,6 | **6** | **8** | **8** | **9** |
| I=6  | 4   | 6   | 6   | 6   | **7** | **8** | **8** | **10** |
| I=7  | 4   | 6   | 7   | 7   | **7** | **9** | **9** | **10** |
| I=8  | 4   | 6   | 8   | 8   | 8,9 | **9** | **10** | **11** |
| I=9  | 4   | 6   | 8   | 9   | 9   | **9** | **10** | **12** |
| I=10 | 4   | 6   | 8   | 9   | 10  | 10  | **10** | **12** |
| I=11 | 4   | 6   | 8   | 9   | 11  | 11  | **11** | **13** |
| I=12 | 4   | 6   | 8   | 9   | 12  | 12  | 12,13 | **13** |

Table 2: Generic rank of unconstrained arrays of dimension $K \times K \times K$.

| K | 2 | 3 | 4 | 5 | 6 | 7 | 8 | 9 |
|---|---|---|---|---|---|---|---|---|
| R | 2 | 5 | **7** | **10** | **14** | **19** | **24** | **30** |

We report values of the generic rank of 3-way arrays with equal dimensions in table 2. Kruskal [12, p.9] refers to a "much studied $9 \times 9 \times 9$ array whose rank has been bounded between 18 and 23 but is still unknown". One can observe that the code `rangj3(K,K,K)` yields a generic rank of $\bar{R} = 30$ for $K = 9$, which shows that the array Kruskal refers to is sub-generic. The values shown in table 2 can also be compared to those obtained in the symmetric case (see table 6).

Now the algorithm can be run on tensors of order higher than 3. To make the presentation of the results readable, table 3 reports values of the generic rank obtained for asymmetric tensors with equal dimensions, $N$, and order $L$, with an algorithm referred to as `rangj(N,L)`. We also indicate

Table 3: (top) Generic rank of unconstrained arrays of equal dimensions, $N$, and order $L$. (bottom) Number of remaining degrees of freedom; essential uniqueness of the CAND occurs when $F = 0$.

| $\bar{R}$ | N=2 | N=3 | N=4 | N=5 | N=6 | N=7 | N=8 |
|---|---|---|---|---|---|---|---|
| $L = 3$ | 2 | 5 | 7 | 10 | 14 | 19 | 24 |
| $L = 4$ | 4 | 9 | 20 | 37 | 62 | | |

| $F$ | N=2 | N=3 | N=4 | N=5 | N=6 | N=7 | N=8 |
|---|---|---|---|---|---|---|---|
| $L = 3$ | 0 | 8 | 6 | 5 | 8 | 18 | 16 |
| $L = 4$ | 4 | 0 | 4 | 4 | 6 | | |

the dimensionality of the fiber of solutions. This number is simply defined as the difference:

$$F(N, L) = \bar{R}(N, L)(LN - L + 1) - N^L$$

Only those values of dimension and order for which $F = 0$ have an essentially unique CAND, that is, unique up to usual scale and permutation ambiguities.

**Tensors with symmetric matrix slices.** Having verified that `rangj3` and `rangj` work correctly throughout the cases were the generic/typical rank are known, we next turn to the $I \times J \times J$ arrays with I symmetric slices (Table 4). Again, known values coincide with numerical ones delivered by the code `rgindscal3`. We inserted results obtained from `rgindscal3` alone in bold face. As far as can be determined, all results are again in agreement with previously known values.

**Tensors with double centered symmetric slices.** When the symmetric slices are also row-wise (or column-wise, which is the same thing) zero-mean, the code `rgindscal2z` yielded the values reported in table 5. Note that the generic rank computed by `rgindscal2z(J,I)` is the same as that computed by `tgindscal3(J-1,I)`, at least according to the values explored in table 4. This confirms the conjecture we made earlier in this paper.

**Symmetric tensors.** Finally, we also report in table 6 values obtained with 3-way or 4-way symmetric arrays, obtained with the code `rangjs`. Note that these results have been already reported in [5]. The dimensionality of the fiber of solutions is given by:

$$F(N, L) = \bar{R} N - \binom{N + L - 1}{L}$$

It is interesting to compare the ranks with those of the unsymmetric case, obviously larger, reported in table 3. In particular, by inspection of the values of $F$, one can observe that uniqueness is again rarely met with generic arrays, but less rarely than in the non-symmeric case.

Table 4: Typical ranks for $I \times J \times J$ arrays, with $J \times J$ symmetric slices. Values reported in bold correspond to generic ranks computed numerically. Values separated by commas are known typical ranks.

|      | J=2 | J=3 | J=4  | J=5 |
|------|-----|-----|------|-----|
| I=2  | 2,3 | 3,4 | 4,5  | 5,6 |
| I=3  | 3   | 4   | **6**    | **7**   |
| I=4  | 3   | 4,5 | **6**    | **8**   |
| I=5  | 3   | 5,6 | **7**    | **9**   |
| I=6  | 3   | 6   | **7**    | **9**   |
| I=7  | 3   | 6   | **7**    | **10**  |
| I=8  | 3   | 6   | **8**    | **10**  |
| I=9  | 3   | 6   | 9,10 | **11**  |
| i=10 | 3   | 6   | 10   | **11**  |

Table 5: Typical ranks for $I \times J \times J$ arrays, with $J \times J$ symmetric slices having zero-mean columns.

|      | J=2 | J=3 | J=4 | J=5 |
|------|-----|-----|-----|-----|
| I=2  | 1   | 2   | 3   | 4   |
| I=3  | 1   | 3   | 4   | 6   |
| I=4  | 1   | 3   | 4   | 6   |
| I=5  | 1   | 3   | 5   | 7   |
| I=6  | 1   | 3   | 6   | 7   |
| I=7  | 1   | 3   | 6   | 7   |
| I=8  | 1   | 3   | 6   | 8   |
| I=9  | 1   | 3   | 6   | 9   |
| i=10 | 1   | 3   | 6   | 10  |

Table 6: (top) Generic ranks of symmetric arrays of dimension $N$ and order $L$. (bottom) Number of remaining degrees of freedom; essential uniqueness of the CAND occurs when $F = 0$.

| $R$   | N=2 | N=3 | N=4 | N=5 | N=6 | N=7 | N=8 |
|-------|-----|-----|-----|-----|-----|-----|-----|
| $L=3$ | 2   | 4   | 5   | 8   | 10  | 12  | 15  |
| $L=4$ | 3   | 6   | 10  | 15  | 21  | 30  | 42  |
| $F$   | N=2 | N=3 | N=4 | N=5 | N=6 | N=7 | N=8 |
| $L=3$ | 0   | 2   | 0   | 5   | 4   | 0   | 0   |
| $L=4$ | 1   | 3   | 5   | 5   | 0   | 0   | 6   |

# Appendix: Computer codes

The codes[1] are provided in this section in the form of simple Scilab functions (Scilab is a softaware made available by Inria, freely to academic laboratories; its syntax ressembles that of Matlab).

```
function [Rbar,D,F]=rangj3(N1,N2,N3)
// [Rbar,D,F]=rangj3(N1,N2,N3)
// P.Comon, 22 May 2006.
// Rbar is the generic rank of tensors of order 3 and dimensions N1xN2xN3,
// and D is the dimension of the whole space of square tensors
// F is the dimension of the variety of solutions
// if a norm constraint is imposed on loading vectors in each term,
// F=F=Rbar*(N1+N2+N3-2)-D
// Call:   getf("rangj3.sci");stacksize(2000000);
// [Rbar,D,F]=rangj3(N1,N2,N3)
// Example: K=4;for I=2:9,for J=3:5,M(I-1,J-2)=rangj3(I,J,K);end;end;M
I1=eye(N1,N1);I2=eye(N2,N2);I3=eye(N3,N3);JAC=[];
 rankold=0;ranknew=1;
 while ranknew>rankold,
  a=rand(1,N1,"normal");b=rand(1,N2,"normal");c=rand(1,N3,"normal");
  JAC=[JAC;kron(I1,kron(b,c));kron(a,kron(I2,c));kron(a,kron(b,I3))];
 temp=ranknew;ranknew=rank(JAC);
 printf('ranknew=%i\n',ranknew)
 rankold=temp;
 end;
 D=ranknew;
```

---

[1]These codes can be downloaded from the URL www.i3s.unice.fr/~pcomon, both in Scilab and Matlab formats.

```
[lign,col]=size(JAC);Rbar=lign/(N1+N2+N3)-1;F=Rbar*(N1+N2+N3-2)-D;

function [Rbar,D,F]=rangj(N,K)
// [Rbar,D]=rangj(N,K)
// P.Comon, 21 May 2006.
// Rbar is the generic rank of square tensors of order K and dimension N,
// and D is the dimension of the whole space of square tensors
// F is the dimension of the variety of solutions
// if a norm constraint is imposed on K-1 loading vectors in each term,
// F=Rbar*(K*N-K+1)-D
// Call:   getf("rangj.sci");stackize(2000000);
// for n=2:5,for k=3:4,[R,D]=rangj(n,k);G(n-1,k-2)=R;F(n-1,k-2)=R*(k*n-k+1)-D;printf('\n');end;end
// NB: this implementation limited to K=3 or K=4, and to N<6 for reasons of
// numerical complexity and memory requirements.
getf("/users/comon/Documents/Scilab/fonctions/gsm.sci")
I=eye(N,N);JAC=[];
if K==3,
 rankold=0;ranknew=1;
 while ranknew>rankold,
  a=rand(1,N,"normal");b=rand(1,N,"normal");c=rand(1,N,"normal");
  JAC=[JAC;kron(I,kron(b,c));kron(a,kron(I,c));kron(a,kron(b,I))];
 temp=ranknew;ranknew=rank(JAC);
 // temp=ranknew;[Q,U]=gsm(JAC);ranknew=rank(U);
 printf('ranknew=%i\n',ranknew)
 rankold=temp;
 end;
 D=ranknew;
elseif K==4,
 rankold=0;ranknew=1;
 while ranknew>rankold,
  a=rand(1,N,"normal");b=rand(1,N,"normal");c=rand(1,N,"normal");d=rand(1,N,"normal");
  JAC=[JAC;kron(I,kron(b,kron(c,d)));kron(a,kron(I,kron(c,d)));kron(a,kron(b,kron(I,d)));kron(a,kr
 temp=ranknew;ranknew=rank(JAC);
 // temp=ranknew;[Q,U]=gsm(JAC);ranknew=rank(U);
 printf('ranknew=%i\n',ranknew)
 rankold=temp;
 end;
 D=ranknew;
else
  Rbar=0;D=0;
end;
```

```
[lign,col]=size(JAC);Rbar=lign/N/K-1;F=Rbar*(K*N-K+1)-D;

function [Rbar,D,F]=rgindscal3(N2,N3)
// [Rbar,D,F]=rgindscal3(N2,N3)
// P.Comon, 29 June 2006.
// Rbar is the generic rank of tensors of order 3 and dimensions (N2,N2,N3)
// with symmetric slices across the two first modes
// and D is the dimension of the whole tensor space
// F is the dimension of the variety of solutions
// if a norm constraint is imposed on K-1 loading vectors in each term,
// F=Rbar*(N2+N3-1)-D
// rmin is a lower bound on the generic rank, if known. The default is 1.
// Call:   getf("rgindscal3.sci");stacksize(2000000);
// for I=2:10, for J=2:5, M(I-1,J-1)=rgindscal3(J,I);end;end;M
I2=eye(N2,N2);I3=eye(N3,N3);JAC=[];
 rankold=0;ranknew=1;
 while ranknew>rankold,
  b=rand(1,N2);c=rand(1,N3);
  JAC=[JAC;kron(I2,kron(b,c))+kron(b,kron(I2,c));kron(b,kron(b,I3))];
 temp=ranknew;ranknew=rank(JAC);
 printf('ranknew=%i\n',ranknew)
 rankold=temp;
 end;
 D=ranknew;
[lign,col]=size(JAC);Rbar=lign/(N2+N3)-1;F=Rbar*(N2+N3-1)-D;

function [Rbar,D,F]=rgindscal2z(N,K)
// [Rbar,D,F]=rgindscal2z(N,K)
// P.Comon, 1st July 2006.
// Rbar is the generic rank of tensors of order 3 and dimensions (N,N,K)
// with symmetric slices across the two first modes, and doubly centered
// and D is the dimension of the whole tensor space
// F is the dimension of the variety of solutions
// if a norm constraint is imposed on loading vectors in each term,
// F=Rbar*(N+K-2)-D
// Call:   getf("rgindscal2z.sci");stacksize(2000000);
// for K=2:15, for N=2:5, M(K-1,N-1)=rgindscal2z(N,K);end;end;M
I3=eye(K,K);JAC=[];IZ=[eye(N-1,N-1),-ones(N-1,1)];
 rankold=0;ranknew=1;
 while ranknew>rankold,
  b=rand(1,N-1);bb=[b,-sum(b)];c=rand(1,K);
```

```
   JAC=[JAC;kron(IZ,kron(bb,c))+kron(bb,kron(IZ,c));kron(bb,kron(bb,I3))];
 temp=ranknew;ranknew=rank(JAC);
 printf('ranknew=%i\n',ranknew)
 rankold=temp;
 end;
 D=ranknew;
[lign,col]=size(JAC);Rbar=lign/(N-1+K)-1;F=Rbar*(N+K-2)-D;

function [Rbar,D]=rangjs(N,d)
// [Rbar,D]=rangjs(N,d)
// Rbar is the generic rank of symmetric tensors of order d and dimension N,
// and D is the dimension of the whole space of symmetric tensors
// F=N*Rbar-D is the dimension of the variety of solutions
// P.Comon, 10 January 1996; modified for scilab 20 may 2006.
// Call: stacksize(2000000);
// getf("rangjs.sci");for i=2:6,for j=3:4,M(i-1,j-2)=rangjs(i,j);end;end;M
// NB: preferably limited to d=3 or d=4 and to N<7 for reasons of memory space
I=eye(N,N); JAC=[];
if d==3,
 rankold=0;ranknew=1;
 while ranknew>rankold,
  a=rand(1,N,"normal");
  JAC=[JAC;kron(I,kron(a,a))+kron(a,kron(I,a))+kron(a,kron(a,I))];
 temp=ranknew;[Q,U]=gsm(JAC);ranknew=rank(U);
 printf('ranknew=%i\n',ranknew)
 rankold=temp;
 end;
 D=ranknew;
elseif d==4,
 rankold=0;ranknew=1;
 while ranknew>rankold,
  a=rand(1,N,"normal");
  JAC=[JAC;kron(I,kron(a,kron(a,a)))+kron(a,kron(I,kron(a,a)))+kron(a,kron(a,kron(I,a)))+kron(a,kr
 temp=ranknew;[Q,U]=gsm(JAC);ranknew=rank(U);
 printf('ranknew=%i\n',ranknew)
 rankold=temp;
 end;
 D=ranknew;
else
 Rbar=0;D=0;
end;
```

```
[lign,col]=size(JAC);Rbar=lign/N-1;
```